\documentstyle[twoside,epsf,psfig]{article}

\catcode`\@=11
\long\def\@makefntext#1{
\protect\noindent \hbox to 3.2pt {\hskip-.9pt  
$^{{\eightrm\@thefnmark}}$\hfil}#1\hfill}		

\def\@makefnmark{\hbox to 0pt{$^{\@thefnmark}$\hss}}	
	
\def\ps@myheadings{\let\@mkboth\@gobbletwo
\def\@oddhead{\hbox{}
\rightmark\hfil\eightrm\thepage}   
\def\@oddfoot{}\def\@evenhead{\eightrm\thepage\hfil
\leftmark\hbox{}}\def\@evenfoot{}
\def\sectionmark##1{}\def\subsectionmark##1{}}

\oddsidemargin=\evensidemargin
\newcounter{sectionc}\newcounter{subsectionc}\newcounter{subsubsectionc}
\renewcommand{\section}[1] {\vspace{12pt}\addtocounter{sectionc}{1} 
\setcounter{subsectionc}{0}\setcounter{subsubsectionc}{0}\noindent 
	{\tenbf\thesectionc. #1}\par\vspace{5pt}}
\renewcommand{\subsection}[1] {\vspace{12pt}\addtocounter{subsectionc}{1} 
\setcounter{subsubsectionc}{0}\noindent 
{\bf\thesectionc.\thesubsectionc. {\kern1pt \bfit #1}}\par\vspace{5pt}}
\renewcommand{\subsubsection}[1] {\vspace{12pt}\addtocounter{subsubsectionc}{1}
	\noindent{\tenrm\thesectionc.\thesubsectionc.\thesubsubsectionc.
	{\kern1pt \tenit #1}}\par\vspace{5pt}}
\newcommand{\nonumsection}[1] {\vspace{12pt}\noindent{\tenbf #1}
	\par\vspace{5pt}}

\newcounter{appendixc}
\newcounter{subappendixc}[appendixc]
\newcounter{subsubappendixc}[subappendixc]
\renewcommand{\thesubappendixc}{\Alph{appendixc}.\arabic{subappendixc}}
\renewcommand{\thesubsubappendixc}
	{\Alph{appendixc}.\arabic{subappendixc}.\arabic{subsubappendixc}}

\renewcommand{\appendix}[1] {\vspace{12pt}
        \refstepcounter{appendixc}
        \setcounter{figure}{0}
        \setcounter{table}{0}
        \setcounter{lemma}{0}
        \setcounter{theorem}{0}
        \setcounter{corollary}{0}
        \setcounter{definition}{0}
        \setcounter{equation}{0}
        \renewcommand{\thefigure}{\Alph{appendixc}.\arabic{figure}}
        \renewcommand{\thetable}{\Alph{appendixc}.\arabic{table}}
        \renewcommand{\theappendixc}{\Alph{appendixc}}
        \renewcommand{\thelemma}{\Alph{appendixc}.\arabic{lemma}}
        \renewcommand{\thetheorem}{\Alph{appendixc}.\arabic{theorem}}
        \renewcommand{\thedefinition}{\Alph{appendixc}.\arabic{definition}}
        \renewcommand{\thecorollary}{\Alph{appendixc}.\arabic{corollary}}
        \renewcommand{\theequation}{\Alph{appendixc}.\arabic{equation}}
        \noindent{\tenbf Appendix \theappendixc #1}\par\vspace{5pt}}
\newcommand{\subappendix}[1] {\vspace{12pt}
        \refstepcounter{subappendixc}
        \noindent{\bf Appendix \thesubappendixc. {\kern1pt \bfit #1}}
	\par\vspace{5pt}}
\newcommand{\subsubappendix}[1] {\vspace{12pt}
        \refstepcounter{subsubappendixc}
        \noindent{\rm Appendix \thesubsubappendixc. {\kern1pt \tenit #1}}
	\par\vspace{5pt}}

\newcommand{\textlineskip}{\baselineskip=13pt}
\newcommand{\smalllineskip}{\baselineskip=10pt}

\newcommand{\copyrightheading}[1]
	{\vspace*{-2.5cm}\smalllineskip{\flushleft
	{\footnotesize International Journal of Modern Physics D, #1}\\
	{\footnotesize \copyright\kern2pt World Scientific Publishing
	 Company}\\
	 }}

\newcommand{\publisher}[2]{{\begin{center}\footnotesize\smalllineskip 
	Received #1\\
	Revised #2
	\end{center}
	}}

\def\abstracts#1#2#3{{
	\centering{\begin{minipage}{4.5in}\footnotesize\baselineskip=10pt
	\parindent=0pt #1\par 
	\parindent=15pt #2\par
	\parindent=15pt #3
	\end{minipage}}\par}} 

\newcommand{\bibit}{\nineit}

\renewenvironment{thebibliography}[1]
        {\frenchspacing
	 \ninerm\baselineskip=11pt
         \begin{list}{\arabic{enumi}.}
        {\usecounter{enumi}\setlength{\parsep}{0pt}     
	 \setlength{\leftmargin 12.7pt}{\rightmargin 0pt}
         \setlength{\itemsep}{0pt} \settowidth
	{\labelwidth}{#1.}\sloppy}}{\end{list}}

\newcounter{itemlistc}
\newcounter{romanlistc}
\newcounter{alphlistc}
\newcounter{arabiclistc}

\newcommand{\fcaption}[1]{
        \refstepcounter{figure}
        \setbox\@tempboxa = \hbox{\footnotesize Fig.~\thefigure. #1}
        \ifdim \wd\@tempboxa > 5in
           {\begin{center}
        \parbox{5in}{\footnotesize\smalllineskip Fig.~\thefigure. #1}
            \end{center}}
        \else
             {\begin{center}
             {\footnotesize Fig.~\thefigure. #1}
              \end{center}}
        \fi}

\newcommand{\tcaption}[1]{
        \refstepcounter{table}
        \setbox\@tempboxa = \hbox{\footnotesize Table~\thetable. #1}
        \ifdim \wd\@tempboxa > 5in
           {\begin{center}
        \parbox{5in}{\footnotesize\smalllineskip Table~\thetable. #1}
            \end{center}}
        \else
             {\begin{center}
             {\footnotesize Table~\thetable. #1}
              \end{center}}
        \fi}

\def\@citex[#1]#2{\if@filesw\immediate\write\@auxout
	{\string\citation{#2}}\fi
\def\@citea{}\@cite{\@for\@citeb:=#2\do
	{\@citea\def\@citea{,}\@ifundefined
	{b@\@citeb}{{\bf ?}\@warning
	{Citation `\@citeb' on page \thepage \space undefined}}
	{\csname b@\@citeb\endcsname}}}{#1}}

\newif\if@cghi
\def\cite{\@cghitrue\@ifnextchar [{\@tempswatrue
	\@citex}{\@tempswafalse\@citex[]}}
\def\citelow{\@cghifalse\@ifnextchar [{\@tempswatrue
	\@citex}{\@tempswafalse\@citex[]}}
\def\@cite#1#2{{$\null^{#1}$\if@tempswa\typeout
	{IJCGA warning: optional citation argument 
	ignored: `#2'} \fi}}

\def\pmb#1{\setbox0=\hbox{#1}
	\kern-.025em\copy0\kern-\wd0
	\kern.05em\copy0\kern-\wd0
	\kern-.025em\raise.0433em\box0}

\def\fnm#1{$^{\mbox{\scriptsize #1}}$}
\def\fnt#1#2{\footnotetext{\kern-.3em
	{$^{\mbox{\scriptsize #1}}$}{#2}}}

\def\runninghead#1#2{\pagestyle{myheadings}
\markboth{{\protect\footnotesize\it{\quad #1}}\hfill}
{\hfill{\protect\footnotesize\it{#2\quad}}}}
\font\tenrm=cmr10
\font\tenit=cmti10 
\font\tenbf=cmbx10
\font\bfit=cmbxti10 at 10pt
\font\ninerm=cmr9
\font\nineit=cmti9

\font\eightrm=cmr8

\def\qed{\hbox{${\vcenter{\vbox{	          
   \hrule height 0.4pt\hbox{\vrule width 0.4pt height 6pt
   \kern5pt\vrule width 0.4pt}\hrule height 0.4pt}}}$}}

\begin{document}
\setlength{\textheight}{7.7truein}    

\runninghead{C. Palomba} {Detectability of gravitational radiation $\ldots$}

\normalsize\textlineskip
\thispagestyle{empty}
\setcounter{page}{1}

\copyrightheading{}		

\vspace*{0.88truein}

\centerline{\bf DETECTABILITY OF GRAVITATIONAL RADIATION FROM}
\vspace*{0.035truein}
\centerline{\bf PROMPT AND DELAYED STAR COLLAPSE TO A BLACK HOLE}
\vspace*{0.37truein}
\centerline{\footnotesize CRISTIANO PALOMBA}
\vspace*{0.015truein}
\centerline{\footnotesize\it Dipartimento di Fisica ``G. Marconi'', Universit\`a di Roma
``La Sapienza''}
\baselineskip=10pt
\centerline{\footnotesize\it and Sezione INFN ROMA1, p.le A. Moro 5}
\baselineskip=10pt
\centerline{\footnotesize\it Roma, I-00185, Italy}
\vspace*{0.225truein}
\publisher{(received date)}{(revised date)}

\vspace*{0.21truein}
\abstracts{We consider the emission of gravitational waves in the two proposed models for the collapse of a massive star to a black hole: the {\em prompt} collapse, in which nearly all the star collapses to a black hole in a dynamical time scale, and the 
{\em delayed} collapse, in which a light black hole, or a neutron star,
which subsequently accretes matter, forms  
due to the fall-back achieving, in the neutron star case,
the critical mass for black hole formation. 
Recent simulations strongly support this last scenario.
We show that, due to the slowness of fall-back, in the {\em delayed}
collapse the main burst of gravitational radiation is emitted 
depending on the parameters, mass and angular momentum, of the initial,
light, black hole. 
We estimate, under different assumptions, the detectability of the
emitted gravitational waves showing
that such kind of collapse is not particularly suited for 
detection by forthcoming interferometric detectors.
Detectors with high sensitivity at frequencies greater than $\sim 4\div 5~ kHz$ would be
better suited for this kind of sources.
We calculate also the final mass distribution function of single black holes.}{}{}



\vspace*{1pt}\textlineskip	
\vspace*{-0.5pt}
\noindent

\section{Introduction}
\noindent
Gravitational collapse of a stellar core to a black hole has been studied 
since many years. Efforts have been also devoted to the calculation of the gravitational radiation emitted in this process. Most studies have been based on a perturbative 
approach,\cite{cun1}$^,$\cite{cun2}$^,$\cite{cun3}$^,$\cite{seid1}$^,$\cite{seid2} others on the numerical solution of the full Einstein equations \cite{stark}$^,$\cite{stark2}. All these papers consider the collapse of a ``naked'' stellar core, described by a ``dust'' of particles or, at 
most, by a politropic equation of state, without taking into account the presence of the outer layers of the star, which are involved in the process. 
In particular, depending on the ratio between the energy released in the final explosion of a massive star and the binding energy of the ejected material, two different kinds of collapse have been outlined \cite{woos}$^,$\cite{kalo}: 
the {\em prompt} collapse, in which a large fraction of the star collapses on 
a dynamical time-scale forming a massive black hole, and the {\em delayed} 
collapse, in which a low mass black hole or, alternatively, a neutron star 
forms at the beginning and later accretes 
matter, due to fall-back. If a neutron star is the initial outcome of the
collapse, fall-back pushes its mass above the critical mass
and the formation of a black hole takes
place. In both cases, this light black hole continues to slowly accrete
matter until the final mass is reached.

In the gravitational wave community the {\em prompt} collapse has been 
considered for a long time as representative of realistic collapse processes. 
The estimated mass (several solar masses) of the first black hole candidates 
(like $Cygnus~ X-1$) has led 
to the idea that black holes should often be born with a "typical" mass of
$\sim 10M_\odot $. This assumption appears no more justified now: we know
that the evolution of massive stars in binary systems (to which all
observed black hole candidates belong) is different from 
that of single stars; in addition, more refined observational techniques
have allowed
to find black hole candidates of few solar masses. In the light of these 
results
and of the recent numerical simulations which we will describe, the {\em prompt} birth of 
such massive black holes ($\sim 10M_\odot $ or more) 
should be considered as a very rare event.

In this paper, we will discuss the {\em delayed}
collapse model, the expected mass distribution of isolated black holes 
and
the detectability of the emitted signal, by forthcoming interferometric
detectors. Our main
aim is to understand how our perspectives of detection change with respect 
to the "naive" {\em prompt} collapse.
The plan of the paper is as follows. In Sec.2 we will shortly describe the 
{\em delayed} collapse scenario. 
In Sec.3 we will estimate the contribution of fall-back to the total 
gravitational emission in the {\em delayed} collapse discussing the
consequences. In Sec.4 we will
discuss the
detectability of the emitted signal by forthcoming interferometric
detectors, and compare with the {\em prompt} collapse 
which could happen for progenitor stars above about $40M_\odot $,  
if stellar winds were much less important in the
evolution of massive stars than it is currently believed. 
In Sec.5 we will derive the theoretical final mass distribution function for isolated 
black holes, using the results of recent simulations of the collapse of
massive single stars. Finally, in
Sec.6 the results and their implications will be discussed.
    
\section{Star Collapse to a Black Hole}
\label{starcoll} 
The fate of a massive star ($m_{prog}>9~M_{\odot}$) is the core collapse with 
the formation of a compact object: a neutron star or a black hole.
Numerical simulations show that 
the actual final product, and also the way in which it is formed, depend on 
many factors, among which the mass and the angular momentum of the progenitor, the explosion energy, the high density matter equation of state ($EOS$)
and also the way in which the physics is implemented (regarding, for instance, 
neutrino physics or angular momentum transport).
Moreover, results cannot be considered conclusive until fully relativistic
3D simulations will be performed.
It has been shown
that, for a wide range of progenitor masses and explosion energies, the
shock cannot expell all the matter outside the collapsing core, so that
part of the helium mantle and heavy elements may slow down below the
escape velocity and be accreted by the just formed neutron star, with a 
timescale of minutes to hours. If the neutron star mass grows above a
critical value, a black hole forms and we have a {\em delayed} collapse. 
On the other hand, if the explosion completely fails, or if it is too
weak, a black hole immediately forms: this is the {\em prompt}
collapse. According
to recent simulations (starting from non-rotating progenitors) by Woosley $\&$ Weaver \cite{woos} and by Fryer $\&$
Kalogera \cite{kalo}, typical collapses are always {\em delayed}, the 
{\em prompt} ones occurring only for very massive progenitor stars ($M>40M_\odot $) if stellar winds are negligible, an assumption which appears rather unlikely.
The $EOS$ of high density matter plays a basic role in the determination of 
neutron stars limiting mass, i.e. the critical mass for black hole
formation, $m_{min}$. 
For conventional EOS the neutron star maximum mass ranges between
$1.7~M_{\odot}$ and $2.2~M_{\odot}$ (e.g. \cite{thor}$^,$
\cite{keil}$^,$\cite{bro1}$^,$\cite{akma}), with a $10\div 20 \%$ increase if rotation is taken into account. 
On the other hand, if $\pi$ or $K$ condensation, or
formation of quark matter, takes place at very high densities
(e.g. 
\cite{bro2}$^,$\cite{gled}$^,$\cite{bro3}$^,$\cite{baum}), 
the $EOS$ is softened and this reduces the maximum mass that the pressure
of degenerate
matter can sustain. In this case, the critical neutron star mass is 
$\sim 1.5~M_{\odot}$. 
The masses of 26 neutron stars, measured in pulsars, are compatible with a gaussian distribution with $\overline{m}=1.35\pm 0.04$, and then are consistent with that previous limit \cite{thorcha}.  
Moreover, the 
lack of evidence for neutron stars with mass near the maximum derived from 
conventional $EOS$ is reinforced by the lack of evidence for a pulsar as a remnant 
of the supernova $ SN1987~A$ whose progenitor had a mass of 
$\sim 18~M_{\odot}$, which should have left behind a remnant
mass of about $1.5~M_{\odot}$
\cite{bro1}$^,$\cite{bro3}. However, no definite conclusion can be still
drawn\cite{zampi1}$^,$\cite{zampi2}.
Given the uncertainties in the neutron stars equation of state, in the 
following we will always refer to two different values of the black hole 
minimum mass: $m_{min}=1.5M_{\odot}$, representative of soft equations of 
state (soft EOS), and 
$m_{min}=2M_{\odot}$ for standard, with no phase transition, equations of
state (conventional EOS).  

\section{Gravitational Radiation from Fall-Back}
In this section, we want to estimate the contribution of fall-back to
the
emission of gravitational waves in the {\em delayed} collapse of a massive star to a black hole. 

There are many studies on the capture by a black hole of finite-size
shells of matter \cite{was1}$^,$\cite{was2}$^,$\cite{sasaki}. 
A general result is that the amount of gravitational radiation emitted is 
always smaller than that emitted by the capture of a pointlike particle
with the same mass of the shell. This is a consequence of destructive 
interference of the radiation emitted by different parts of the
infalling
extended matter. These results are confirmed also by the recent, more 
realistic, calculations by Papadopoulos $\&$ Font \cite{pappa}. In the case of 
an axisymmetric irrotational shell of matter with mass $\mu$, much smaller
than 
the mass $m$ of the black hole, 
they find that the efficiency in the emission of gravitational waves 
decreases as a function of the radial width $L$ of the shell as follows:
\begin{equation}
\epsilon_{s} ={\Delta E\over{mc^2}}=8\cdot 10^{-3}\left({\mu\over
m}\right)^2
\left({m\over{M_{\odot}}}\right)^{2.4}\left({L\over {1~km}}\right)^{-2.4}
\label{effishell}
\end{equation}
We shall now extrapolate this equation for $\mu > m$ and compare to the
efficiency
we expect from core collapse
which, for a maximally rotating core, is $\epsilon_{c}\sim 7\cdot 10^{-4}$
in the axisymmeric case \cite{stark}. 
For instance, assuming that the initial mass of the newly formed black
hole is 
$m=2~M_{\odot}$, and that the mass of the shell is
$\mu=10~M_{\odot}$, we
find that the condition $\epsilon_s<0.1\cdot  \epsilon_c$ requires
$L>100~km$. This condition is largely verified for massive stars.  
Then, we can conclude that the main burst of gravitational radiation is emitted at the formation of 
the black hole while subsequent accretion of matter gives no important 
contributions. To this respect, two possibilities can occur.
First, the $Fe$ core of the star has a mass greater than $m_{min}$.
In such a case the black hole initial mass is equal to the core mass. 
Second, the $Fe$ core 
is lighter than $m_{min}$.
In this case, a neutron star is produced at the beginning.
It then accretes matter until the critical mass is reached, so that
the black hole initial mass is equal to the minimum one. 
From the core masses given by Woosley $\&$ Weaver \cite{woos}
we see that for soft EOS ($m_{min}=1.5~M_{\odot}$) all stars with initial
mass greater than
$18~M_{\odot}$ produce cores that immediately collapse to a black hole of mass greater
than $m_{min}$, but lower than $\sim 1.8~M_{\odot}$. The same happens according to Fryer \cite{fryer}. 
For conventional EOS ($m_{min}=2~M_{\odot}$) black holes are formed,
after fall-back on a
neutron star, from the collapse of progenitors of mass $m_{prog}>26~M_{\odot}$ (following Fryer $\&$ Kalogera \cite{kalo} this value is reduced to 
$m_{prog}\sim 20~M_\odot$). 
Then, 
all black holes form with an initial mass equal, or 
nearly equal, to the minimum one, presumably in the range $1.5\div
2.6~M_{\odot}$\fnm{e}\fnt{e}{The upper limit takes 
into account also the possible stabilizing effect of rotation.}, depending on 
the nuclear density matter
equation of state and on the rotation rate. 
In the following we will assume, for simplicity of calculation, that all black holes form with the same mass $m_{min}$, so that their initial mass distribution 
function, for the {\em delayed} collapse scenario, can be written as a 
$\delta$-function: $f(m)= \delta(m-m_{min})$.
We will consider $m_{min}=1.5~M_{\odot}$ and $m_{min}=2~M_{\odot}$. 

\subsection{The rate of black hole formation}
Let us now estimate the expected event rate for collapses. We assume a 
Galactic rate for Supernovae of
type $II$ given by $R_{SNII}=0.02~yr^{-1}$. Such rate is the sum of the rate 
of collapses to neutron stars  
and, in the {\em delayed} scenario, of the rate of collapses to black hole (a supernova explosion is
produced by all {\em delayed} collapses, indipendently of $m_{min}$). 
For instance, for soft EOS, we have seen that a black hole is 
produced by progenitor stars with mass $18M_\odot \le m_{prog} \le 
40M_\odot $. Their formation rate, $R_{bh}$,
is a fraction $\lambda $ of
the rate of formation of neutron stars $R_{ns}$,
that are generated by progenitors with mass $9M_{\odot}\le m_{prog} \le
18M_{\odot}$. Of course, this ratio depends on the initial mass function
of the progenitor stars: $\lambda=0.43$ for a Salpeter law with 
exponent $\alpha=-2.35$, while $\lambda=0.32$ for a Scalo law with 
$\alpha=-2.7$. Then, we can write 
\begin{equation}
R_{SNII}=R_{ns}+R_{bh}=R_{ns}(1+\lambda),
\label{rsn}
\end{equation}
from which we can get $R_{ns}$ and $R_{bh}$.       
In Tab.(\ref{rate}) we 
tabulate the galactic rates of black hole formation in the different 
cases, considering both {\em delayed} ($R_{bh}$) and {\em prompt} 
($R_{bh,p}$) collapses. Moreover, the total expected rate within 
the Virgo cluster, $R_{@20Mpc}$, is given in the last column. 
\begin{table}$$
\begin{tabular}{|c|c|c|c|} \hline
$W\&W~m_{prog}/m~relation$&$ R_{bh}~[yr^{-1}] $&$ R_{bh,p}~[yr^{-1}] $&$
R_{@20Mpc}~[yr^{-1}]     
$\\ \hline
$ Salpeter,~~m_{min}=1.5~M_{\odot} $&$ 8.5\cdot 10^{-3} $&$ 
3.4\cdot 10^{-3} $&$ 5.1~ (2.0) $\\ \hline
$ Salpeter,~~m_{min}=2~M_\odot $&$ 2.8\cdot 10^{-3} $&$ 2.7\cdot 10^{-3}
$&$ 1.7~ (1.6)  $\\ \hline
$ Scalo,~~m_{min}=1.5~M_\odot $&$ 6.6\cdot 10^{-3} $&$ 1.9\cdot 10^{-3} 
$&$ 3.8~ (1.1) $\\ \hline
$ Scalo,~~m_{min}=2~M_{\odot} $&$ 1.9\cdot 10^{-3} $&$ 1.5\cdot 10^{-3} 
$&$ 1.2~ (0.9) $\\ \hline
$F\&K~m_{prog}/m~relation$&$~$&$~$&$~$\\ \hline
$ Salpeter,~~m_{min}=1.5~M_{\odot} $&$ 8.5\cdot 10^{-3} $&$ 
3.4\cdot 10^{-3} $&$ 5.1~ (2.0)  $\\ \hline
$ Salpeter,~~m_{min}=2~M_\odot $&$ 6.3\cdot 10^{-3} $&$ 3.1\cdot 10^{-3} 
$&$ 3.8 ~(1.9)  $\\ \hline
$ Scalo,~~m_{min}=1.5~M_\odot $&$ 6.6\cdot 10^{-3} $&$ 1.9\cdot 10^{-3} 
$&$ 3.8~ (1.1) $\\ \hline
$ Scalo,~~m_{min}=2~M_{\odot} $&$ 4.8\cdot 10^{-3} $&$ 1.8\cdot 10^{-3} 
$&$ 2.9~ (1.1) $\\ \hline
\end{tabular}$$
\caption{Galactic event rate for {\em delayed} black 
hole  formation ($R_{bh}$), assuming a galactic type-$II$ Supernova rate 
$R_{SNII}=0.02~ yr^{-1}$, using 
Woosley $\&$ Weaver$^{13}$ ($W\& W$) and Fryer $\&$ Kalogera$^{44}$ ($F\&
K$) progenitor mass-remnant mass relations. 
``Salpeter'' means that an 
exponent $\alpha=-2.35$ has been used in the progenitor mass distribution 
law, while ``Scalo'' 
corresponds to $\alpha=-2.7$. The galactic rate for {\em prompt} collapses, 
which could occur for $m_{prog}\in [40,120]M_\odot$, is also given
($R_{bh,p}$). In the last 
column, the total expected rate for
{\em delayed} collapses within $20~Mpc$, $R_{@20Mpc}$, is indicated 
(in parenthesis the corresponding values for {\em prompt} collapses).}
\label{rate}
\end{table}
From Tab.(\ref{rate}) we see that the fraction of {\em delayed} collapses
producing a black hole is in the range $10\% \div 43\%$ of those leading to a 
neutron star. These percentages increase to $18\% \div 60\% $ if {\em
prompt} collapse, for stars more massive than $40~M_{\odot}$, is also
taken into account. 

\section{The Detectability}
Gravitational signals emitted in star collapse to a 
black hole are characterized by an initial part, emitted during the in-fall 
phase and bounce, followed by an oscillating tail, which can be described
as a superposition of damped
sinusoids, corresponding to the black hole quasi-normal modes. The frequency 
and the damping time of the quasi-normal modes are a function of the black 
hole parameters, mass and angular momentum. For axisymmetric collapses, 
the main contribution
(typically $\sim 90\%$ of the whole energy emitted) is given by the $l=2$ mode,
for which the frequencies and the damping times are given, as a function
of the rotation parameter $a=J/{\left({Gm^2\over c}\right)}$, in
Tab.(\ref{qnm}). 
\begin{table}$$
\begin{tabular}{|c|c|c|} \hline
$ a $&$n\cdot \nu~[kHz] $&$ \tau /n~[10^{-5}s] $\\ \hline
$ 0.0 $&$ 12.0556 $&$ 5.5472 $ \\ \hline
$ 0.2 $&$ 12.1007 $&$ 5.5660 $ \\ \hline
$ 0.4 $&$ 12.2491 $&$ 5.6230 $ \\ \hline
$ 0.6 $&$ 12.5201 $&$ 5.7407 $ \\ \hline
$ 0.8 $&$ 12.9653 $&$ 6.0061 $ \\ \hline
$ 0.9 $&$ 13.2911 $&$ 6.2892 $ \\ \hline
$ 0.98 $&$ 13.6234 $&$ 6.7170 $ \\ \hline
$ 0.9998 $&$ 13.7137 $&$ 6.8760 $ \\ \hline
\end{tabular}$$
\caption{Frequencies and damping times of the first $l=2$, $m=0$ quasi-normal 
mode of a 
rotating black hole as a function of the rotation parameter 
$a=J/{\left({Gm^2\over c}\right)}$. 
The mass of the black hole is $m=nM_\odot $. From Ferrari $\&$
Palomba$^{31}$.}
\label{qnm}
\end{table}
We use the waveforms calculated by Stark $\&$ Piran 
\cite{stark}, who have computed, in the framework of the full non-linear
theory, the gravitational signals emitted in the axisymmetric collapse to a 
black hole of a rotating core, for different values of the angular momentum. 
The maximum efficiency reached in their simulations is $\epsilon_{max}\simeq 7\cdot 10^{-4}$.
From the waveforms we have computed the corresponding one-sided power spectrum $f(\nu)$, 
i.e. the flux of energy per unit frequency which is given by
\begin{equation}
f(\nu)={\pi c^3 \nu^2\over{2G}}<h^2(\nu)>
\label{fnu}
\end{equation}
where $<h^2>$ denotes the average of the squared gravitational signal with respect to its angular dependence.
These computed energy spectra allow to evaluate
the signal to noise ratio ($SNR$), according to the well-known
formula given by Eq.(\ref{snrav}) below,
which holds if the matched filter is applied to the data.
This optimum filtering procedures implies that we know the exact waveform emitted and this 
is a rather optimistic hypothesis even in the case of star collapses to a black hole. We 
know that in the simplest cases (low rotation, no hydrodynamical effects) most of the energy 
is emitted in the phase of quasi-normal ringing, which can
be described as a superposition of damped sinusoids. On the other hand, if 
the collapse has 
a high degree of rotation, or if strong hydrodynamical effects take place,
the emitted waveform 
may be no longer dominated by the quasi-normal ringing and would be less
predictable \cite{ferpal}. 
In this case, other, less optimum, data analysis procedures should be used
\cite{hallo}. Then, 
our results have to be considered as upper limits on the $SNR$. 
The average squared $SNR$ can be expressed as
\begin{equation}
\overline{SNR^2}={8G\over {5\pi c^3}}
\int_0^{\infty}{f(\nu)\over {\nu^2 S_h(\nu)}}d\nu~ .
\label{snrav}
\end{equation}
where $S_h(\nu)$ is the 
detector noise power spectrum. 
In Eq.(\ref{snrav}) an average on the source-detector relative position and on the 
polarization of the gravitational waves emitted is also performed. 
As discussed in Sec.3, the {\em delayed} collapse 
produces black holes that, immediately after birth, have mass very close to 
the 
minimum mass. Thus, the emitted signal depends essentially on the black hole 
angular momentum and on the distance at which the collapse takes place. 
In {\em prompt} collapses, on the contrary, the $SNR$ clearly depends on the mass of the 
black hole and, 
due to the typical sensitivity curve of ground-based interferometers, the
radiation emitted by more 
massive black holes is more easily detectable, so that the {\em delayed} collapse process is 
definitely less favourable than the ``naive'' {\em prompt} one. 
In Tab.(\ref{tab1}) we give the $SNR=\sqrt{\overline{SNR^2}}$ for {\em delayed} collapses taking place in the Galaxy 
(we fix $r=10kpc$), for different values of the minimum black hole mass and of the rotation 
parameter. 
\begin{table}$$
\begin{tabular}{|c|c|} \hline
$~~$&$ {SNR} $\\ \hline
$ a=0.42 $&$  $ \\ \hline
$ m_{min}=1.5M_{\odot} $&$ {6.1} $ \\ \hline
$ m_{min}=2M_{\odot} $&$ {7.9} $ \\ \hline
$a=0.79 $&$  $ \\ \hline
$ m_{min}=1.5M_{\odot} $&$ {25.4} $ \\ \hline
$ m_{min}=2M_{\odot} $&$ {31.9} $ \\ \hline
$ a=0.94 $&$  $ \\ \hline
$ m_{min}=1.5M_{\odot} $&$ {42.1} $ \\ \hline
$ m_{min}=2M_{\odot} $&$ {51.4} $ \\ \hline 
\end{tabular}$$
\caption{$Virgo$ detector: signal-to-noise ratio for the {\em delayed} collapse. Different 
values of the rotation parameter $a$ and minimum mass for black hole formation $m_{min}$ are 
considered.
All collapses are assumed to take place at a distance $r=10kpc$.}
\label{tab1}
\end{table}
The values of the $SNR$ increase with the angular 
momentum of the black hole. This is the consequence of two effects: first, the efficiency of 
emission increases as $a^4$; second, for 
very high angular momentum (say, $a>0.8$) the bounce
of the collapsing 
star produces a lower frequency component in the signal energy spectrum which 
fits better to the sensitivity curve
of interferometers. On the other hand, the expected rate of detectable events 
is low: collapses to a black hole are detectable, essentially, only 
within the Local Group, 
with a total rate, strongly dominated by the Milky Way, which is less than 
$\sim 1$ event 
per century, see Tab.(\ref{rate}).    
We stress, however, that higher $SNR$ could be obtained if collapses had a 
higher degree of 
asymmetry and there are some observative indications supporting this
hypothesis
\cite{strom}$^,$\cite{wang}$^,$\cite{tayl}$^,$\cite{reed}.
Moreover, the initial core collapse to a neutron star which takes 
place, in the {\em delayed} case, if $m_{min}=2~M_\odot $, could be a
promising process. If it is highly asymmetric, as observations seem to
indicate, 
a large amount of gravitational radiation could be emitted in the range of frequencies where 
interferometric detectors reach their best 
sensitivity\fnm{f}\fnt{f}{It should be noted, however, that according to
the newtonian 
simulations by Rampp {\em et al.} \cite{rampp}, the amount of gravitational radiation 
emitted in non-axisymmetric collapses is comparable to that of the axisymmetric case.}.
 
\begin{figure}[htbp] 
\vspace*{13pt}
\centerline{\psfig{file=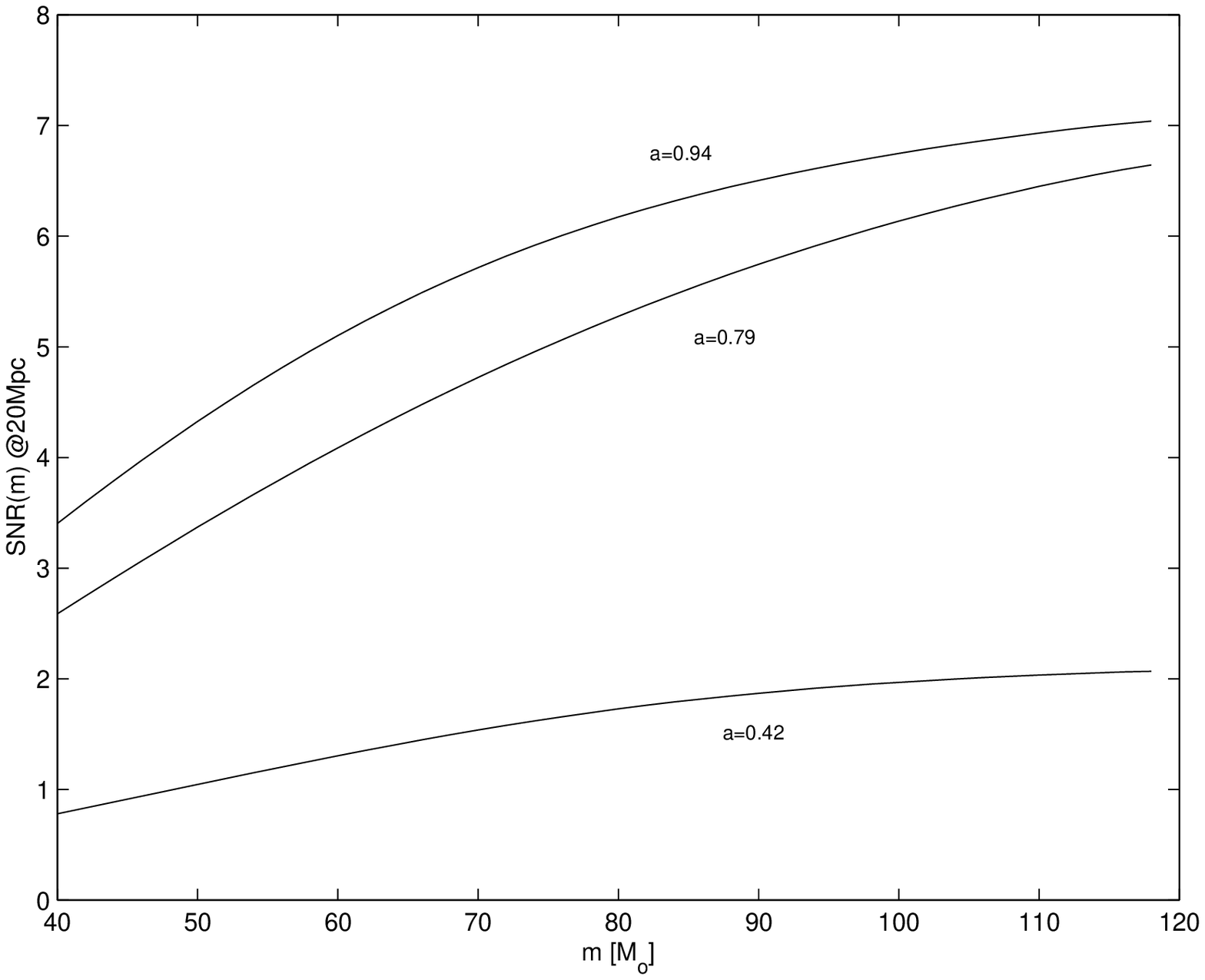,width=8cm,height=5cm}} 
\vspace*{13pt}
\fcaption{Advanced $LIGO$ detector: signal-to-noise ratio as a function of the black hole mass $m$ for {\em prompt} collapses. The collapse is assumed to take place at a distance $r=20~Mpc$.}
\label{snrm}
\end{figure}

We have repeated the 
calculation of the $SNR$ for the advanced $LIGO$ detector, using an approximation to its 
sensitivity curve given by Flanagan $\&$ Huges \cite{fla}.
\begin{table}$$
\begin{tabular}{|c|c|} \hline
$~~$&$ {SNR} $\\ \hline
$ a=0.42 $&$  $ \\ \hline
$ m_{min}=1.5M_{\odot} $&$ {0.013} $ \\ \hline
$ m_{min}=2M_{\odot} $&$ {0.019} $ \\ \hline
$a=0.79 $&$  $ \\ \hline
$ m_{min}=1.5M_{\odot} $&$ {0.065} $ \\ \hline
$ m_{min}=2M_{\odot} $&$ {0.088} $ \\ \hline
$ a=0.94 $&$  $ \\ \hline
$ m_{min}=1.5M_{\odot} $&$ {0.12} $ \\ \hline
$ m_{min}=2M_{\odot} $&$ {0.16} $ \\ \hline 
\end{tabular}$$
\caption{Advanced $LIGO$ detector: signal-to-noise ratio for the {\em delayed} collapse. 
Different values of the rotation parameter $a$ and minimum mass for black hole formation 
$m_{min}$ are considered.
All collapses are assumed to take place at a distance $r=20Mpc$.}
\label{tab3}
\end{table}
Results are given, for {\em delayed} collapses, in Tab.(\ref{tab3}) where a distance of 
$r=20Mpc$ has been assumed. Within this distance, corresponding
approximately to the Virgo Cluster, the 
expected black hole formation rate is $\sim 1\div 5~yr^{-1}$, 
but we see that the detection perspectives 
are not much better because the $SNR$ is much smaller than one. The
situation is different for {\em prompt} collapses. 
In Fig.(\ref{snrm}) we have plotted the signal-to-noise ratio for {\em prompt} collapses as 
a function of the newborn black hole mass and it appears that the emitted gravitational radiation could be detectable if the rotation rate is high enough.   

\section{Black hole final mass distribution function}
In this section we calculate the black hole final mass distribution
function. 
We do not consider the chemical evolution of the galaxy, which affects the
mass distribution of compact remnants (see, e.g., Timmes {\em et al.}
\cite{timm}). Such a theoretical distribution could be useful in statistical studies
of recently formed black holes.

\begin{figure}[htbp] 
\vspace*{13pt}
\centerline{\psfig{file=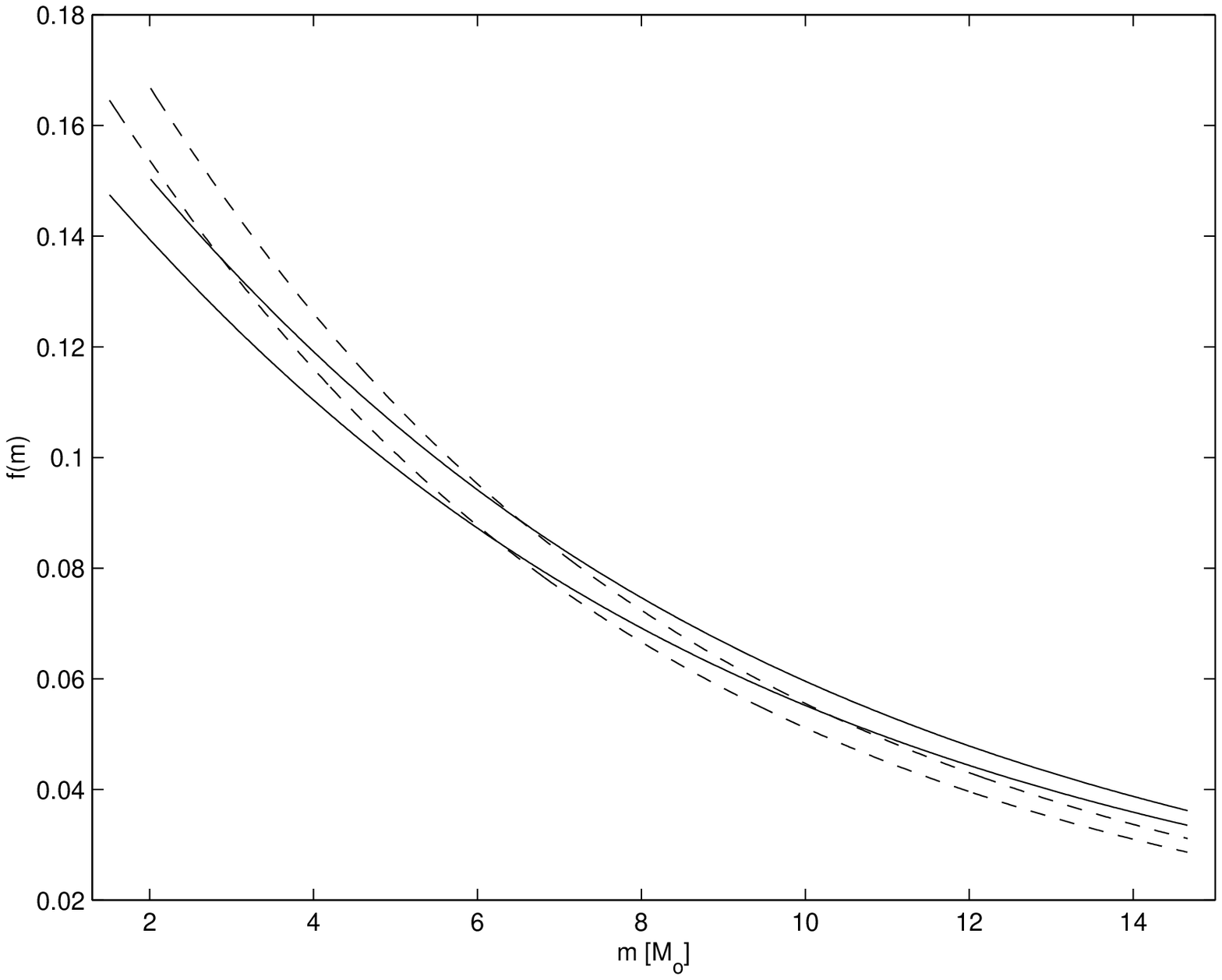,width=8.0cm,height=5cm}} 
\vspace*{13pt}
\fcaption{Black hole final mass distribution functions, obtained from the results of 
Fryer $\&$ Kalogera \cite{kalo}, for $m_{min}=1.5~M_{\odot}$ and $m_{min}=2~M_{\odot}$. 
Solid lines refer to a progenitor mass distribution function with exponent 
$\alpha=-2.35$, while dashed lines are for $\alpha=-2.7$.}
\label{fmzsolall_kalo}
\end{figure}

The black hole final mass distribution depends on the mass distribution of 
progenitor stars and on the relation between the mass of progenitors and that 
of the final remnants produced after the collapse. For the first one we assume 
a power law 
$f(m_{prog})\propto m^{\alpha}_{prog}$,
with $\alpha =-2.35$ (Salpeter's law) and $\alpha =-2.7$ (Scalo's law). Regarding the relation between the mass of progenitor stars, $m_{prog}$, and that of remnants, $m$, we use (after conversion from baryonic to gravitational mass) the results of Woosley $\&$ Weaver \cite{woos} 
and of Fryer $\&$ Kalogera \cite{kalo}, who have performed systematic calculations of the evolution and explosion of non rotating massive stars. It must be stressed again that such evolutionary calculations are still subject to many uncertainties and 
this reflects in some differences between their results, obtained using
different assumptions.  
The remnant masses they report are calculated a long time
after the collapse, so that the possible fall-back is included. 
The distribution function we are searching for is simply given by
\begin{equation}
f(m)=f(m_{prog}(m))\cdot {dm_{prog}\over {dm}}
\label{mbh}
\end{equation}
\begin{figure}[t] 
\vspace*{13pt}
\centerline{\psfig{file=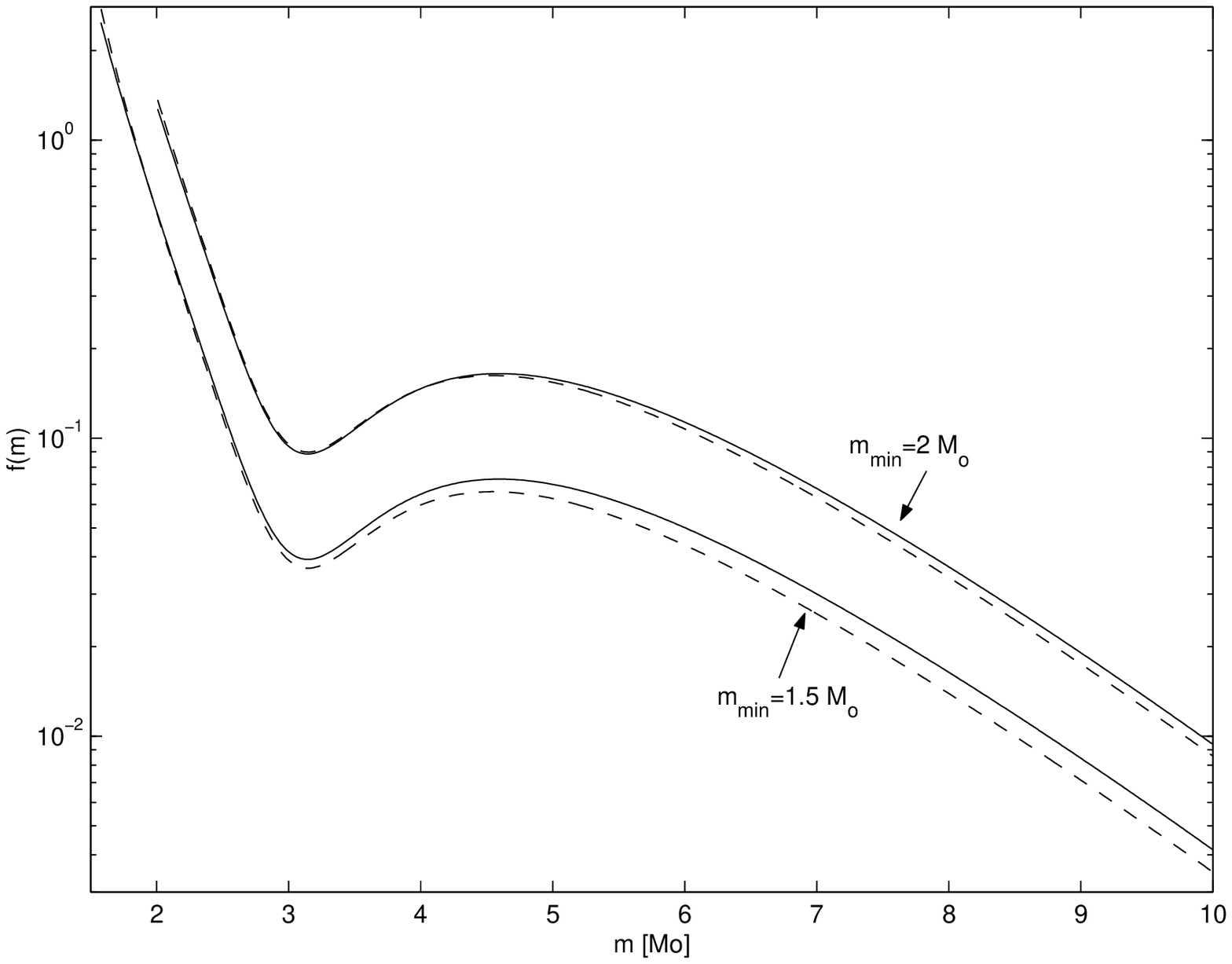,width=8.0cm,height=5cm}} 
\vspace*{13pt}
\fcaption{Black hole final mass distribution function, obtained from the results of 
Woosley $\&$ Weaver \cite{woos}, for $m_{min}=1.5~M_{\odot}$ and $m_{min}=2~M_{\odot}$. 
Solid lines refer to a progenitor mass distribution function with exponent 
$\alpha=-2.35$, while dashed lines are for $\alpha=-2.7$.}
\label{fmzsolall}
\end{figure}
In Fig.(\ref{fmzsolall}) we plot the function $f(m)$ 
obtained using a fit of the progenitor mass-remnant mass relation found by 
Woosley $\&$ Weaver \cite{woos} (see their table 3), considering different
values of
$m_{min}$ and of the exponent $\alpha$. 
The maximum progenitor mass considered by Woosley $\&$ Weaver is 
$m_{prog}=40~M_{\odot}$. According to them, for masses greater than this,
the effect of
stellar wind becomes relevant and 
strongly modifies the star evolutionary path: 
a smooth convergence of remnant masses to 
$\sim 1.5~M_{\odot}$, corresponding to neutron stars or low mass black
holes,
is expected for progenitors of mass greater than about $40~M_{\odot}$.
As a consequence, black holes with mass greater than $\sim 10.3~M_\odot$ - which is what 
is predicted by Woosley $\&$ Weaver for a $40~M_\odot$ progenitor 
star if stellar winds are neglected - should never be formed, for
progenitors of solar 
metallicity.\fnm{a}\fnt{a}{Black holes of mass greater than that value
could 
be produced if the metallicity is lower.}
Also Fryer $\&$ Kalogera \cite{kalo} show that low mass remnants are
produced, for progenitors more massive than about $40~M_\odot$,
if stellar winds are taken into account.  
On the other hand, if stellar winds are neglected, they find that, for
masses above about $40~M_\odot$,
the star collapse takes place in the {\em prompt} way, i.e. a black hole of 
mass nearly equal to the mass of the progenitor star is produced in a
dynamical timescale. In this case the black hole mass distribution
function is simply proportional to $m^{-\alpha}$.
The final mass distribution function,
resulting from the progenitor mass-remnant mass 
relation found by Fryer $\&$ Kalogera, is plotted in 
Fig.(\ref{fmzsolall_kalo}) 
(we refer to their ``most likely'' model, in which $50\%$ of the explosion 
energy goes in unbinding the star). 
There are clear differences between the distributions functions
plotted in Figs.(\ref{fmzsolall},\ref{fmzsolall_kalo}).
First of all, the expected range of masses for black holes is larger 
in the second case, reaching $\sim 14.7M_\odot $. Moreover, in this case the 
function $f(m)$ is monotonically decreasing, while in
Fig.(\ref{fmzsolall}) there is a minimum around $3~M_{\odot}$
followed by a relative maximum at $\sim 4.5~M_{\odot}$. These differences
are a 
consequence, obviously, of the different relations for progenitor
{\em vs.} remnant masses\fnm{b}\fnt{b}{In particular, the minimum in
Fig.(\ref{fmzsolall}) is due to an increase of the slope of $m_{prog}(m)$
for $m_{prog}\simeq 30~M_{\odot}$ while the following maximum is produced
by the first term in the right-hand side of Eq.(\ref{mbh}), which is
dominant for large $m_{prog}$.} and are due, mainly, to
two reasons. 
First, explosion simulations are carried on by Woosley $\&$ Weaver in 
one-dimension, while Fryer $\&$ Kalogera simulations are
two-dimensional. Second,
Woosley $\&$ Weaver constrain the kinetic energy of ejected material at 
infinity\fnm{c}\fnt{c}{Defined as the difference 
between the explosion energy and the binding energy of the ejecta} to a constant value, about $1.2\cdot 
10^{51}erg $, while in Fryer $\&$ Kalogera this quantity decreases for increasing progenitor mass. This explains why their remnant masses are greater.     

The theoretical distribution functions we have found cannot be compared with observations, which refer to black candidates belonging to binary systems. It is widely accepted that the evolution of high mass stars in binary systems is quite different from that of single stars of 
comparable mass. A detailed discussion 
on massive
binary systems evolution and on the possible mechanisms and bias effects 
which can explain the 
observed black hole candidates mass distribution can be found in refs. 
\cite{fryer} and \cite{palo}.  

\section{Conclusions}
In this paper we have considered the collapse of a massive star to a black hole, exploring 
the models of {\em prompt} and {\em delayed} collapse. According to recent simulations, the 
formation of stellar mass black holes should take place through a {\em delayed} collapse. 
The {\em prompt} collapse could happen for very massive progenitors (mass greater than 
$\sim 40M_\odot$) only if stellar winds were negligible, an assumption which appears to be 
not very reliable.   
In the {\em delayed} collapse all black holes were born with a mass equal
to its minimum 
value, or just a little greater, and then slowly accrete matter up to
their final mass. We have 
shown that the main burst of gravitational radiation is emitted when the black hole forms so 
that the gravitational energy spectrum is peaked in the range 
$\sim 4.5\div 9~kHz$, depending 
on the initial mass and the angular momentum of the black hole. Such frequencies do not match very well with 
the sensitivity curve of ground-based interferometers which reach their
best sensitivity in 
the band $\sim 60\div 1000~Hz$.
We have estimated the detectability of the
emitted gravitational 
radiation  in the {\em delayed} case, the most reliable scenario, and in the 
{\em prompt} case, for progenitor stars more massive than $40M_\odot $. For 
each 
collapse model, we have considered different values of black hole minimum mass and angular 
momentum and different laws for progenitors mass distribution.  
We have also derived the theoretical black hole final mass distribution
function.

{\em Delayed} collapses are detectable only inside the Local Group of
galaxies by 
interferometers of the first generation. Obviously, {\em delayed} collapses are less 
detectable than the {\em prompt} ones, due to the lower matching of their
characteristic 
frequencies to the sensitivity curve of the detectors. On the other hand, the initial stage 
of the {\em delayed} collapse, with the formation of a neutron star following 
a strongly asymmetric 
explosion (if $m_{min}=2~M_\odot $), could be a promising source of gravitational waves. 
Detection perspectives of {\em delayed} collapses are not so much better for 
adavnced interferometers because, if 
distances up to the Virgo Cluster are considered (where we expect $\sim
1\div 5~ev/yr$), the 
signal-to-noise ratio is much lower than one, unless a very high degree of 
asimmetry is produced. On the contrary, {\em
prompt} collapses could be detected with large enough $SNR$. 
{\em Delayed} collapses to a black hole belong to a class of high frequency sources of 
gravitational waves, which comprises various processes involving compact 
objects, as for instance the excitation of neutron star w-modes
\cite{kokko},
the coalescence of two neutron stars with the formation of a light black 
hole \cite{baum2},
dynamical instabilities in neutron stars \cite{centr},
and some kinds of secular instability in neutron stars \cite{lai}.
Such sources cannot be efficiently detected by present resonant detectors and 
forthcoming 
interferometers because they are expected to 
emit at frequencies higher than those at which both interferometric and 
resonant detectors have their best sensitivity. 
In past years {\em local arrays} of small
resonant detectors, which are particularly suited for the detection of
high frequency gravitational radiation, have been
proposed\cite{frasca1}$^,$\cite{frasca2}. A detailed study of their 
detection performances, considering different geometries, dimensions and 
materials, has been done \cite{palo}.  
  
The non-continuous background of gravitational waves produced by the ensemble of the star 
collapses to a black hole, which occurred at a higher rate 
in the early 
phases of the Universe, has recently been calculated \cite{fersch}. 
It would be interesting to repeat the calculation in the case of {\em
delayed} collapse.
In such a case we have the superposition of energy spectra nearly with the same shape and 
all peaked at nearly the same frequency, in the range $4.5\div 9kHz$ (this holds also for 
star metallicity $Z=0$). 
Roughly speaking, as most of the collapses take place at redshift 
$z\sim 2$, we expect to have, at the detector, a background spectrum strongly peaked 
somewhere in the band $1.5\div 3~kHz$. 

\nonumsection{Aknowledgements}
I want to thank S. Frasca and M. A. Papa for the useful discussions 
and suggestions, T. Piran for his encouragement, V. Ferrari for her careful reading of the 
manuscript and the anonymous referees for 
their comments aiming to an improvement of this paper.

\nonumsection{References}

\end{document}